\documentclass[11pt,english,nofootinbib]{revtex4}
\usepackage[T1]{fontenc}
\usepackage[latin9]{inputenc}
\usepackage{amstext}
\usepackage{amssymb}
\usepackage{esint}
\usepackage{babel}
\usepackage{color}
\def\be{\begin{equation}}
\def\ee{\end{equation}}
\def\nn{\nonumber}
\begin{document}

\title{Quantum gravity with torsion and non-metricity}

\author{C. Pagani} 
\affiliation{Institut f\"{u}r Physik (WA THEP) Johannes-Gutenberg-Universit\"{a}t, Staudingerweg 7, 55099 Mainz, Germany}
\author{R.~Percacci} 
\affiliation{International School for Advanced Studies, 
via Bonomea 265, 34136 Trieste, Italy} 
\affiliation{INFN, Sezione di Trieste, Italy}

\begin{abstract}
We study the renormalization of theories of gravity with an arbitrary
(torsionful and non-metric) connection.
The class of actions we consider is of the Palatini type,
including the most general terms with up to two derivatives of the metric, 
but no derivatives of the connection. 
It contains 19 independent parameters.
We calculate the one loop beta functions of these parameters and find
their fixed points.
The Holst subspace is discussed in some detail and found not to be stable under renormalization.
Some possible implications for ultraviolet and infrared gravity are discussed.
\end{abstract}
\maketitle


\section{Extended theories of gravity}

There are several ways of formulating the equations of Einstein's General Relativity:
they differ in the type and number of fields that are initially present, 
but  the set of solutions (at least for pure gravity, not coupled to matter fields)
is always the same.
There is thus motivation to investigate the quantum properties of these off-shell extensions, 
in the hope that some of them may be better behaved than others.
One important class of generalizations of GR consists of
theories with independent connection. Usually this means
connections with torsion, but more generally one could also
have connections that are not metric-compatible.
In this paper we shall discuss the renormalization of
a class of theories of this type.

Scalar and gauge fields are differential forms and their Lagrangians do not require a gravitational connection.
However, spinor fields carry representations of the Lorentz
or orthogonal group and therefore can only couple 
to metric connections.
Thus, they could act as sources of torsion \cite{hehl,shapiroReview}.
We will see, however, that the existence of spinor fields, by itself,
does not imply that the gravitational connection must be metric:
the spinor could couple to a metric connection constructed with the
dynamical connection and the metric.

The standard way of formulating General Relativity is in terms of a dynamical metric $g_{\mu\nu}$.
In this case one has ten fields, and the action is invariant under the group of diffeomorphisms,
which is parametrized by four ``gauge functions''.
The tetrad formulation, which is used to write the coupling of spinors to gravity,
uses sixteen dynamical fields $\theta^a{}_\mu$, related to the metric by
\be
\label{tetra}
g_{\mu\nu}=\theta^a{}_\mu\theta^b{}_\nu \eta_{ab}\ .
\ee
The additional six fields are rendered unphysical by local Lorentz invariance,
which is parameterized by exactly six gauge functions.
One can go one step further in this direction and extend the local gauge invariance to the
sixteen-dimensional general linear group $GL(4)$
\cite{perbook,flop,higgsgr,cuiaba,siegel,leclerc,kirsch}.
One then has two fields: an ``internal'' metric $\gamma_{ab}$ and a ``soldering form''
$\theta^a{}_\mu$, related to the metric by 
\be
\label{gl4}
g_{\mu\nu}=\theta^a{}_\mu\theta^b{}_\nu \gamma_{ab}\ .
\ee
The general linear group acts on $\gamma_{ab}$ by similarity transformations
and on $\theta^a{}_\mu$ by left multiplication, so that either one of these fields
can be brought to the standard form $\theta^a{}_\mu=\delta^a_\mu$ or $\gamma_{ab}=\eta_{ab}$
by a $GL(4)$ transformation, but not both simultaneously.
In this way the $GL(4)$-invariant formulation reduces to the metric or to the
vierbein formulation by making different gauge choices.
Given a certain gravitational action $S(g_{\mu\nu})$ in the metric formulation,
one can define in a unique way an action in the tetrad formulation $S'(\theta^a{}_\mu)
=S(g_{\mu\nu}(\theta))$ where $g(\theta)$ is given by (\ref{tetra})
and an action in the $GL(4)$ formulations 
$S''(\gamma_{ab},\theta^a{}_\mu)=S(g_{\mu\nu}(\gamma,\theta))$,
where $g(\theta,\gamma)$ is given by (\ref{gl4}).
Still further generalizations are possible
\cite{lippoldt}.
These actions are dynamically equivalent.
There is therefore a well-defined sense in which these different formulations 
can be said to be equivalent {\it for any action}.

A different class of extensions consists of treating the connection as an independent variable.
In Einstein's original formulation of the theory, the connection is identified with
the Levi-Civita connection, which is the unique connection that is metric and torsion-free.
The Levi-Civita connection takes different forms depending on the formulation one is working with.
In the metric formulation, its components are given by the Christoffel symbols:
\be
\mit\Gamma_{\alpha\beta\gamma}={1\over 2}\left(\partial_\alpha g_{\beta\gamma}
+\partial_\gamma g_{\beta\alpha}-\partial_\beta g_{\alpha\gamma}\right)\ ,
\ee
in the vierbein formulation it is given by:
\be
\mit\Gamma_{abc}=-{1\over 2}\left(f_{abc}+f_{cab}-f_{bca}\right)\ .
\ee
and in the $GL(4)$-invariant formulation it is given by:
\be
\mit\Gamma_{abc}={1\over 2}\left(E_{abc}+E_{cab}-E_{bca}\right)
-{1\over 2}\left(f_{abc}+f_{cab}-f_{bca}\right)\ .
\ee
where
\footnote{Latin indices are raised and lowered with $\gamma_{ab}$,
greek indices are raised and lowered with $g_{\mu\nu}$;
greek indices are transformed to latin, and vice-versa, with $\theta^a{}_\mu$
and $\theta_a{}^\mu=(\theta^{-1})_a{}^\mu$.}
\be
E_{abc}=\theta_a{}^\mu\partial_\mu\gamma_{bc}\ ;\qquad
f_{ab}{}^c=(\theta_a{}^\mu\partial_\mu\theta_b{}^\nu
-\theta_b{}^\mu\partial_\mu\theta_a{}^\nu)\theta^c{}_\nu\ .
\ee
In any case, it can be thought of as a composite field.
The most general linear connection in the tangent bundle, $A_\mu{}^a{}_b$, 
will have both torsion and nonmetricity. 
The physical meaning of these quantities is clearest in the $GL(4)$-formalism,
where they are defined by
\be
\label{torsion}
T_\mu{}^a{}_\nu=
\partial_\mu\theta^a_\nu-
\partial_\nu\theta^a_\mu
+A_\mu{}^a{}_b\,\theta^b_\nu
-A_\nu{}^a{}_b\,\theta^b_\mu\ 
\ee
and
\be
\label{nonmetricity}
-Q_{\mu ab}=
\partial_\mu\gamma_{ab}
-A_\mu{}^c{}_a\gamma_{cb}-A_\mu{}^c{}_b\gamma_{ac}\ .
\ee
They are therefore the covariant derivatives of the fields $\gamma_{ab}$ and $\theta^a{}_\mu$,
which in a certain sense can be viewed as Goldstone bosons.
The form of these tensors in the metric and tetrad formulations are obtained
by simply setting $\theta^a{}_\mu=\delta^a_\mu$ or $\gamma_{ab}=\eta_{ab}$.
Notice that $T_\mu{}^\alpha{}_\nu$ involves no derivatives in the metric formulation
and $Q_{a\mu\nu}$ involves no derivatives in the tetrad formulation.
This is a manifestation of a kind of Higgs phenomenon, whereby the
gravitational connection becomes massive \cite{higgsgr,cuiaba}.

The present work is devoted to these theories of gravity with generic
independent connection.
In general these theories will have many propagating degrees of freedom
in addition to a massless spin two graviton \cite{sezgin}.
In contrast to the extensions discussed above,
there is no natural way of identifying an action for the metric and connection 
with an action written for the metric alone,
so these extensions depend very much on the choice of action.
There is however a subclass of theories that are equivalent to Einstein's theory:
it has actions that involve at most two derivatives of the metric
(or tetrad, or soldering form, or internal metric) and no derivatives of the connection.
Indeed such an action will necessarily consist of a combinations of the following:
a cosmological term, a term linear in the Ricci scalar of $A_\mu{}^a{}_b$, 
terms quadratic in $T$ and $Q$.
As we shall review in section 3, the equations of motion of such actions
require that the connection be equal to the Levi-Civita connection,
and therefore such theories are generically equivalent to the
Hilbert action (possibly with cosmological term) on shell.
This is a generalization of the Palatini formalism of GR.

The reason why these particular actions have such properties
can be better appreciated using the logic of effective field theories.
Einstein's theory is singled out among all possible generally covariant
theories constructed with a metric by having equations of motion that
involve at most two derivatives.
These are the terms that dominate the dynamics at distance scales large
relative to the Planck length, and so it is only natural that they should
describe well the gravitational dynamics of large bodies.
The generalized Palatini actions can be seen as gauge theories
for the linear group where the connection (or more precisely the
difference between the dynamical connection and the Levi-Civita connection)
is massive due to the occurrence of a Higgs phenomenon.
This explains why the connection is the
Levi-Civita connection at low energies.
Naively one would expect the mass to be of the order of the Planck mass,
but classically this need not be the case and it is an interesting question to
ask whether some components of the connection may have lower masses,
perhaps low enough to become accessible to accelerators.
This is one of the motivations of the present work.

Another motivation, as already mentioned earlier, is the issue of the
the UV behavior of different off-shell extensions of Einstein's theory.
There is mounting evidence that ``Quantum Einstein Gravity''
(by which we mean a theory of gravity based on the metric as the field carrying
of the degrees of freedom, but not necessarily with the Hilbert action)
may have a fixed point with finitely many UV attractive directions,
thus giving rise to an UV complete and predictive quantum field theory of gravity.
There have already been some attempts to extend this result to
the tetrad formulation \cite{Harst:2012ni,Dona:2012am} 
and to the case with independent connection
\cite{DaumReuter1,BenedettiSpeziale,DaumReuter2,shapiroteixeira}.
These have been limited, however, to metric connections, and then only
to the so-called Holst action,
which contains only one specific combination of torsion squared terms.

In this paper we consider the renormalization of the most general Palatini-like action containing squares of torsion and nonmetricity.
The analysis will be limited to one loop.
The main results are as follows.
First we consider pure gravity with torsion and nonmetricity,
in the absence of matter.
When we consider the off-shell beta functions for the cosmological
constant, Newton's constant and the quadratic torsion and nonmetricity
couplings, in addition to the Gaussian fixed point
there are fixed points where the first two are nonzero
while the torsion and nometricity couplings are either 
zero or infinity.
This agrees with and generalizes previous results where only
certain torsion terms had been taken into account
\cite{DaumReuter1,BenedettiSpeziale,DaumReuter2}.
Furthermore, when Dirac fields are coupled to torsion,
there is also another fixed point where 
also the torsion couplings are finite.

One could take this as evidence that asymptotic safety
persists also in the larger theory space where the connection
is treated as an independent variable.
However, we find this to be rather weak evidence. 
The generalized Palatini truncation 
amounts to taking into account only mass terms
for the dynamical connection.
This is certainly a good approximation at low energy,
where by ``low'' we mean lower than the mass itself,
but it is too crude for a proper analysis of the UV limit.
Above the Planck scale the connection is expected to
have new propagating degrees of freedom,
for which terms quadratic in curvature are needed.
A proper analysis of the UV limit of this teory would have to
take into account such terms.
If we restrict ourselves to ``low'' energy,
in the sense defined above,
and if we view this as an effective field theory, 
then the generic conclusion that can be drawn from
our analysis is that generically all components of the connection
are expected to have Planck-scale mass. 
Much lower scales would require unnatural tunings.

\section{General connections: non-metricity and torsion}

For the sake of simplicity, in this paper we will work in the metric formalism.
We refer to \cite{perbook,higgsgr,cuiaba} for the discussion of similar theories in the $GL(4)$-invariant formalism.
The results of this section are valid in any dimension.

We will denote $A_\mu{}^\rho{}_\nu$ a generic connection in the tangent bundle.
Given a metric $g_{\mu\nu}$, it can be uniquely decomposed into
\be
\label{phi}
A_{\alpha\beta\gamma}=\mit\Gamma_{\alpha\beta\gamma}+\phi_{\alpha\beta\gamma}
\ee
where $\mit\Gamma_{\alpha\beta\gamma}$ is the Levi-Civita connection of $g_{\mu\nu}$
and $\phi_{\alpha\beta\gamma}$ is a tensor without any symmetry properties.
Indices are raised and lowered with $g_{\mu\nu}$.
From (\ref{torsion}) and (\ref{nonmetricity}) one finds
\be
\label{TQphi}
T_{\alpha\beta\gamma} = \phi_{\alpha\beta\gamma}-\phi_{\gamma\beta\alpha}\ ;\qquad
Q_{\alpha\beta\gamma} = \phi_{\alpha\beta\gamma}+\phi_{\alpha\gamma\beta}\ .
\ee
Furthermore $\phi_{\alpha\beta\gamma}$ can be decomposed uniquely into 
\be
\label{alphabeta}
\phi_{\alpha\beta\gamma}= \alpha_{\alpha\beta\gamma}+\beta_{\alpha\beta\gamma}\ ,
\ee
where $\alpha_{\alpha\beta\gamma}$ is symmetric in $(\alpha,\gamma)$
and $\beta_{\alpha\beta\gamma}$ (called the contortion) 
is antisymmetric in $(\beta,\gamma)$:
\begin{eqnarray}
\alpha_{\alpha\beta\gamma} & = & \frac{1}{2}\left(Q_{\alpha\beta\gamma}+Q_{\gamma\beta\alpha}-Q_{\beta\alpha\gamma}\right)\ ,\\
\beta_{\alpha\beta\gamma} & = & \frac{1}{2}\left(T_{\alpha\beta\gamma}+T_{\beta\alpha\gamma}-T_{\alpha\gamma\beta}\right)\ .
\end{eqnarray}
Notice that (\ref{TQphi}) can then also be written as
\be
\label{TQalphabeta}
T_{\alpha\beta\gamma} =  \beta_{\alpha\beta\gamma}-\beta_{\gamma\beta\alpha}\ ,
\qquad
Q_{\alpha\beta\gamma} = \alpha_{\alpha\beta\gamma}+\alpha_{\alpha\gamma\beta}\ ;
\ee
so $\alpha$ contains all the nonmetricity and $\beta$ contains all the torsion.
Another way of saying this is that $\mit\Gamma+\alpha$ is torsion-free
and $\mit\Gamma+\beta$ is metric.
 
We denote $F_{\mu\nu}{}^\rho{}_\sigma$ the curvature tensor of $A_\mu{}^\rho{}_\sigma$,
and $R_{\mu\nu}{}^\rho{}_\sigma$ the curvature tensor of $\mit\Gamma_\mu{}^\rho{}_\sigma$.
They are related as follows:
\begin{eqnarray}
F_{\mu\nu}{}^\alpha{}_\beta & = & 
R_{\mu\nu}{}^\alpha{}_\beta
+\nabla_{\mu}\phi_{\nu \,\,\, \beta}^{\,\,\, \alpha}-\nabla_{\nu}\phi_{\mu\,\,\, \beta}^{\,\,\, \alpha}
+\phi_{\mu\,\,\, \gamma}^{\,\,\, \alpha}\phi_{\nu \,\,\, \beta}^{\,\,\, \gamma}-\phi_{\nu \,\,\, \gamma}^{\,\,\, \alpha}\phi_{\mu \,\,\, \beta}^{\,\,\, \gamma}.
\label{eq:relCurvatures}
\end{eqnarray}
The analog of the Ricci scalar for the connection $A_\mu{}^\alpha{}_\beta$
is the unique contraction $F_{\mu\nu}{}^{\mu\nu}$,
which, up to total derivatives, can be written as
\begin{eqnarray*}
F_{\mu\nu}{}^{\mu\nu} & = & R
+\phi_{\mu \,\,\,\, \gamma}^{\,\,\,\, \mu}\phi_{\nu}^{\,\,\,\, \gamma\nu}-\phi_{\nu\mu\gamma}\phi^{\mu\gamma\nu}\ .
\end{eqnarray*}
This can be reexpressed in terms of non-metricity and
torsion as
\begin{eqnarray}
F_{\mu\nu}{}^{\mu\nu} & = & R
+\frac{1}{4}T_{\alpha\beta\gamma}T^{\alpha\beta\gamma}
 +\frac{1}{2}T_{\alpha\beta\gamma}T^{\alpha\gamma\beta}
 +T_{\alpha}^{\; \alpha\beta}T_{\beta\;\, \gamma}^{\; \gamma}
\nonumber\\
 &  & 
+\frac{1}{4}Q_{\alpha\beta\gamma}Q^{\alpha\beta\gamma}-\frac{1}{2}Q_{\alpha\beta\gamma}Q^{\beta\alpha\gamma}
-\frac{1}{4}Q_{\alpha\gamma}^{\quad \gamma}Q_{\quad \beta}^{\alpha\beta}
+\frac{1}{2}Q_{\alpha}^{\; \alpha\beta}Q_{\beta\gamma}^{\quad \gamma}\nonumber \\
 &  & 
 -Q_{\alpha\beta\gamma}T^{\alpha\beta\gamma}+Q_{\quad \beta}^{\alpha\beta}T_{\alpha \; \gamma}^{\; \gamma}
 -Q_{\alpha}^{\; \alpha \beta}T_{\beta \;\, \gamma}^{\; \gamma}
\label{palatiniTQ}
\end{eqnarray}
or in terms of $\alpha$ and $\beta$ as
\begin{eqnarray}
F_{\mu\nu}{}^{\mu\nu} & = & 
R+\beta_{\,\,\,\, \alpha\beta}^{\alpha}\beta_{\gamma}^{\,\,\,\, \beta\gamma}
+\beta_{\alpha\beta\gamma}\beta^{\beta\alpha\gamma}
-\alpha_{\alpha\beta\gamma}\alpha^{\alpha\gamma\beta}
+\alpha_{\,\,\,\, \alpha\beta}^{\alpha}\alpha_{\gamma}^{\,\,\,\, \beta\gamma}
\nonumber\\
&&
+\alpha_{\alpha\beta\gamma}\beta^{\alpha\beta\gamma}
+\alpha_{\,\,\,\, \alpha\beta}^{\alpha}\beta_{\gamma}^{\,\,\,\, \beta\gamma}
-\alpha_{\,\,\,\, \beta\alpha}^{\alpha}\beta_{\gamma}^{\,\,\,\, \beta\gamma}\ .
\label{palatinialphabeta}
\end{eqnarray}
In the above relations all the possible parity even combinations
of $T$ and $Q$ are generated except $Q_{\alpha}^{\,\,\,\, \alpha\beta}Q_{\,\,\,\, \gamma\beta}^{\gamma}$ or in terms of $\alpha$ and
$\beta$ the following terms are missing: 
$\alpha_{\alpha\beta\gamma}\alpha^{\alpha\beta\gamma}$,
$\alpha_{\alpha}^{\,\,\,\, \beta\alpha}\alpha_{\,\,\,\, \beta\gamma}^{\gamma}$, $\alpha_{\alpha}^{\,\,\,\, \alpha\beta}\alpha_{\beta\gamma}^{\,\,\,\,\,\, \gamma}$
and $\beta_{\alpha\beta\gamma}\beta^{\beta\alpha\gamma}$.

\section{Ansatz}

The (Euclidean) Palatini action for an independent connection is the integral of the only scalar 
that is linear in curvature, plus a possible cosmological term: 
\def\newt{\bar\kappa}
\be
\label{palatini}
S_P(g)=\newt\int d^4x\sqrt{g}(2\Lambda-F_{\mu\nu}{}^{\mu\nu})\ .
\ee
where $\newt=\frac{1}{16\pi G}$.
Using (\ref{palatiniTQ}) or (\ref{palatinialphabeta}),
this can be rewritten as the Hilbert action
\be
\label{hilbert}
S_H(g)=\newt\int d^4x\sqrt{g}(2\Lambda-R)
\ee
plus a specific combination of terms quadratic in torsion and nonmetricity.
There is no reason to restrict our attention to this particular combination
of terms, so we will consider an action that is the sum of 
the Palatini or Hilbert action and the most general term quadratic in torsion and nonmetricity, including also CP-violating terms \cite{hojman}
\be
\label{act}
S=S_H+S_2
\ee
where
\begin{eqnarray}
S_2& = & \int d^4x\sqrt{g}\Big[
a_{1}T_{\lambda \mu 	\nu}T^{\lambda \mu \nu}
+a_{2}T_{\lambda \mu \nu}T^{\lambda \nu \mu}
+a_{3}T_{\: \,\, \mu \lambda}^{\mu}T_{\nu}^{\: \,\, \nu \lambda}\nonumber \\
& &
+b_{1}Q_{\lambda \mu \nu}Q^{\lambda \mu \nu}
+b_{2}Q_{\lambda \mu \nu}Q^{\nu \mu \lambda}
+b_{3}Q_{\lambda \mu}^{\quad \mu}Q_{\quad \nu}^{\lambda \nu}
+b_{4}Q_{\;\,  \mu \lambda}^{\mu}Q_{\nu}^{\;\, \nu \lambda}
+b_{5}Q_{\;\, \mu \lambda}^{\mu}Q_{\,\,\,\,\,\, \nu}^{\lambda \nu}\nonumber \\
&  & 
+c_{1}T_{\lambda \mu \nu}Q^{\lambda \mu \nu}
+c_{2}T_{\;\, \mu \lambda}^{\mu}Q_{\quad \nu}^{\lambda \nu}
+c_{3}T_{\;\, \mu \lambda}^{\mu}Q_{\nu}^{\;\, \nu \lambda}\nonumber \\
&  & +\varepsilon^{\alpha \beta \gamma \delta }g^{\eta \theta}\Bigr(d_{1}T_{\alpha \beta \eta}T_{ \gamma \delta \theta}
+d_{4}T_{\alpha \eta \beta}T_{\gamma \theta \delta}
+d_{5}Q_{\alpha \beta \eta}Q_{\gamma \delta \theta}\nonumber \\
&  & 
\qquad\qquad
+d_{6}Q_{\alpha \eta \theta}T_{\beta \gamma \delta}
+d_{7}Q_{\eta \theta \alpha}T_{\beta\gamma\delta}
+d_{9}Q_{\alpha \eta \beta}T_{\gamma \theta \delta}\Bigr)\Big] \,.
\label{actTQ}
\end{eqnarray}
The nomenclature of the coefficients $d_k$, with $d_2$, $d_3$ and $d_8$ absent,
is of historical origin. 
In fact in intermediate steps of calculations it is necessary to
use an overcomplete set of operators where these coefficients are also present.
We discuss this and other basis choices in detail in Appendix A.
The redundant operators and couplings can be eliminated 
from the final results
and we will not need them in the main text.
We also note here that the first three lines
can be immediately generalized to any dimension,
and only the last two lines are specific to four dimensions.

From an algebraic point of view, the tensors $T$ (antisymmetric in first and third index)
and $Q$ (symmetric in second and third index),
and the tensors $\alpha$ (symmetric in first and third index) 
and $\beta$ (antisymmetric in second and third index)
provide two equally good ways of decomposing a third rank tensor.
However the definitions (\ref{torsion}) and (\ref{nonmetricity})
show that the tensors $T$ and $Q$ are better thought of as covariant derivatives of fields.
Thus we find it more appropriate and, as we shall see, also more convenient from the point of view of calculations,
to treat $\alpha$ and $\beta$ as independent fields.
We therefore reexpress the action $S_2$ as:
\begin{eqnarray}
S_2(\alpha,\beta)& = & 
\int d^4x\sqrt{g}\Bigl[
g_{1}\beta_{\lambda \mu \nu}\beta^{\lambda \mu \nu}
+g_{2}\beta_{\lambda \mu \nu}\beta^{\mu \lambda \nu}
+g_{3}\beta_{\: \lambda \mu}^{\lambda} \beta^\nu{}_\nu{}^\mu 
\nn\\
&  & +g_{4}\alpha_{\lambda \mu \nu}\alpha^{\lambda \mu \nu}
+g_{5}\alpha_{\lambda \mu \nu}\alpha^{\lambda \nu \mu}
+g_{6}\alpha_{\;\, \lambda \mu}^{\lambda}\alpha_{\,\,\,\, n}^{\mu \nu}
+g_{7}\alpha_{\;\, \lambda \mu}^{\lambda}\alpha_{\,\,\,\, n}^{\nu \mu}
+g_{8}\alpha_{\;\, \lambda \mu}^{\mu}\alpha_{\nu}^{\;\, \lambda \nu}
\nn\\
&&
+g_{9}\alpha_{\lambda \mu \nu}\beta^{\lambda \mu \nu}
+g_{10}\alpha_{\;\, \lambda \mu}^{\lambda} \beta^\nu{}_\nu{}^\mu 
+g_{11}\alpha_{\;\, \mu \lambda}^{\lambda} \beta^\nu{}_\nu{}^\mu 
\nn\\
&& 
+\varepsilon_{\alpha \beta \gamma \delta} 
\bigl(
g_{13} \beta^{\alpha \beta \rho} \beta^{\gamma \delta}{}_\rho
+g_{14} \beta^{\rho \alpha \beta} \beta_\rho{}^{\gamma \delta}  
+g_{16} \alpha^{\rho \alpha \beta} \alpha_{\rho}^{\,\,\, \gamma \delta}
\nn\\
&& 
+g_{17}\alpha^{\rho \alpha \beta }\beta_{\rho}^{\,\,\, \gamma \delta}
+g_{19}\alpha_{\,\,\,\, \rho}^{\rho \alpha }\beta^{\beta \gamma \delta}
+g_{20}\alpha_{\,\,\, \rho}^{\rho \,\,\,\, \alpha}\beta^{\beta \gamma \delta}
\bigr)\Bigr].
\label{actalphabeta}
\end{eqnarray}
This action deserves a few comments.
Again one may notice the absence of couplings $d_{12}$, $d_{15}$ and $d_{18}$,
which is due to the existence of relations in an extended basis
containing twenty invariants.
This, and the relation between the couplings in the actions
(\ref{actTQ}) and (\ref{actalphabeta}), is given in Appendix A.

Then, we note that without loss of generality in (\ref{act})
we could replace $S_H$ by $S_P$.
Because of (\ref{palatiniTQ}) and (\ref{palatinialphabeta}),
this would amount just to a shift of the coupligs in $S_2$.
It is however much simpler to work with $S_H$.
For example if one used an action consisting of the Palatini term
plus $S_2$ written in the form (\ref{actTQ}),
there are derivatives acting on the connection
in the Palatini term, derivatives acting on the metric in the $Q$-terms
if one uses coordinate bases
or derivatives acting on the tetrad in the $T$-terms if one used
the tetrad formalism
(or both, is one uses a general linear basis).
This makes for a rather complicated Hessian.
In contrast, we see that the action (\ref{act}) with
$S_2$ written in the form (\ref{actalphabeta})
only contains derivatives in the Hilbert term.
All the rest amounts just to a ``generalized mass term'' for the fields
$\alpha$ and $\beta$
(by which we mean any term quadratic in the fields,
also those involving the $\varepsilon$-tensor).

\section{Beta functions}

A very general and elegant way of defining, and calculating, beta functions,
is to introduce an infrared cutoff in the definition of the effective action,
which is then called the ``Effective Average Action'' or EAA,
to compute the cutoff derivative of the EAA and to extract from it the coefficient
of the desired operator. (For example, the beta function of $1/G$ is the
coefficient of $\int\sqrt{g}R$ in the derivative of the EAA with respect to the cutoff).
This infrared cutoff is implemented by adding to the action a new term
$\Delta S_{k}$, called cutoff action, which suppresses the integration of momentum modes
below a certain scale $k$, i.e.: $p^{2}<k^{2}$.
This cutoff action is quadratic in the fields $\chi$
with a kernel $R_{k}$ depending on momentum: 
$\Delta S_{k}=\frac{1}{2}\int\chi R_{k}\chi$.
One advantage of this process is that one automatically obtains finite quantities,
and that the EAA satisfies a simple equation \cite{wetterich}.
Using the background field method, one can also preserve background gauge invariance.
For these reasons this technique is particularly useful in the case of gravity \cite{reuter1,doupercacci},
where it has been used extensively in the development of the asymptotic safety programme
\cite{
Niedermaier:2006ns,Percacci:2007sz,Litim:2011cp,Reuter:2012id}.

The theory we discuss here is an extension of the 
so-called Einstein-Hilbert truncation.
The main observation which simplifies our task is that in the theory described by (\ref{act}) 
the fields $\alpha_{\mu\nu\rho}$ and $\beta_{\mu\nu\rho}$ do not propagate:
their action is merely a mass term.
We will take this as a sufficient reason not put a cutoff on those degrees of freedom.
\footnote{In principle one could cutoff also nonpropagating degrees of freedom
but we will see that in the subcases that have been analyzed earlier, the result 
of the two procedures agree within theoretical uncertainties such as scheme and gauge choice.}

Since only the graviton propagates, the calculation of the beta functions
of $\Lambda$ and $G$ proceeds exactly as in the familiar Einstein-Hilbert truncation. 
In $d$ dimensions, the one loop beta functions of the dimensionless couplings $\tilde{\Lambda} \equiv \Lambda k^{-2}$ and $\tilde{G} \equiv G k^{d-2}$
read
\begin{eqnarray}
\label{generallambda}
\frac{d\tilde\Lambda}{dt}&=&\!
-2\tilde\Lambda
+\frac{1}{2}A\tilde{G}
-B\tilde G\tilde\Lambda
\\
\label{generalg}
\frac{d\tilde G}{dt}&=&\!\!
(d-2)\tilde G
-B\tilde{G}^2\ .
\end{eqnarray}
The coefficients $A$ and $B$ are non-universal,
but $B$ is always nonzero and positive.
Some calculations involving different cutoff procedures
can be found for example in section IVA in \cite{cpr2}.
The beta functions of the remaining couplings can be found as follows.
First we note that the Hessian has the form:
$S_H^{(2)}+S_2^{(2)}$,
where the first term comes from the Hilbert
part of the ansatz while the second one from the terms having torsion
and non-metricity. To compute the running couplings in (\ref{actalphabeta})
we expand the flow equation in 
$\alpha$ and $\beta$:
\begin{eqnarray*}
\partial_{t}\Gamma_{k} & = & \frac{1}{2}\mbox{Tr}\left[\frac{\partial_{t}R_{k}}{S_H^{(2)}+R_{k}}\right]
-\frac{1}{2}\mbox{Tr}
\left[\frac{1}{S_H^{(2)}+R_{k}}S_2^{(2)}
\frac{1}{S_H^{(2)}+R_{k}}\partial_{t}R_{k}\right]+\cdots.
\end{eqnarray*}
We extract the beta functions from the second term in the above expression
and find a contribution of the following form:
\begin{eqnarray*}
\partial_{t}\Gamma_{k} & \sim & 
-\frac{1}{\newt}Q_{d/2}
\left(\frac{\partial_t R_{k}}{(P_{k}-2\Lambda)^2}\right)
\mbox{Tr}\left({\bf K}^{-1}S_2^{(2)}
\right)\ ,
\end{eqnarray*}
where ${\bf K}$ denotes the tensor
\begin{equation}
 K^{\mu\nu}{}_{\rho\sigma} =
\frac{1}{2} \left( \frac{\delta^\mu_\rho\delta^\nu_\sigma+\delta^\nu_\rho\delta^\mu_\sigma}{2}
+ \frac{1}{2} g^{\mu\nu}g_{\rho\sigma}
\right) \nonumber
\end{equation}
while
\begin{equation}
Q_{n}(f)=\frac{1}{\Gamma[n]}\int_0^\infty dz z^{n-1}f(z)  \nonumber
\end{equation}
(see e.g. \cite{cpr2} for details).
The r.h.s. contains terms proportional to the operators $I_j$ listed in (\ref{basisI}).
The terms proportional to the operators $I_{12}$, $I_{15}$ and $I_{18}$ 
have to be reexpressed in terms of the remaining independent operators
using the identities (\ref{identitiesI}).
Then one reads off the beta functions of the couplings $g_{i}$.
In terms of their dimensionless versions $\tilde g_i=g_i/k^{d-2}$ we find
\begin{eqnarray*}
\partial_{t}\tilde{g}_{1} & = & -\left(d-2\right)\tilde{g}_{1}+\kappa\frac{1}{4} ((d-7) d-12) \tilde{g}_1\\
\partial_{t}\tilde{g}_{2} & = & -\left(d-2\right)\tilde{g}_{2}+\kappa\frac{1}{4} (d-4) (d+1) \tilde{g}_2 \\
\partial_{t}\tilde{g}_{3} & = & -\left(d-2\right)\tilde{g}_{3}+\kappa\frac{1}{4} (d-4) (d+1) \tilde{g}_3\\
\partial_{t}\tilde{g}_{4} & = & -\left(d-2\right)\tilde{g}_{4}+\kappa\left(\frac{1}{4} ((d-7) d-12) \tilde{g}_4 -2 \tilde{g}_8\right)\\
\partial_{t}\tilde{g}_{5} & = & -\left(d-2\right)\tilde{g}_5
+\kappa\left(\frac{1}{4}(d-4)(d+1)\tilde{g}_5+2\tilde{g}_4 \right)\\
\partial_{t}\tilde{g}_{6} & = & -\left(d-2\right)\tilde{g}_{6}
+\kappa\left(\frac{1}{4}(d-4)(d+1)\tilde{g}_6+2 \tilde{g}_4 \right)\\
\partial_{t}\tilde{g}_{7} & = & -\left(d-2\right)\tilde{g}_{7}
+\kappa\left(\frac{1}{4}(d-4)(d+1)\tilde{g}_7+4\tilde{g}_8 \right)\\
\partial_{t}\tilde{g}_{8} & = & -\left(d-2\right)\tilde{g}_{8}
+\kappa\left(\frac{1}{4}(d-8)(d+1)\tilde{g}_8-\tilde{g}_4\right)\\
\partial_{t}\tilde{g}_{9} & = & -\left(d-2\right)\tilde{g}_{9}
+\kappa\frac{1}{4}((d-7)d-16)\tilde{g}_9\\
\partial_{t}\tilde{g}_{10} & = & -\left(d-2\right)\tilde{g}_{10}
+\kappa\left(\frac{1}{4}(d-4)(d+1)\tilde{g}_{10}+\tilde{g}_9 \right)\\
\partial_{t}\tilde{g}_{11} & = & -\left(d-2\right)\tilde{g}_{11}
+\kappa\left(\frac{1}{4}(d-4)(d+1)\tilde{g}_{11}+\tilde{g}_9\right)
\end{eqnarray*}
where
\begin{eqnarray*}
\kappa & = & \frac{16\pi \tilde{G}}{\left(4\pi\right)^{d/2}}
\frac{1}{\left(1-2 \tilde{\Lambda} \right)^{2}}
\frac{2}{\left(d/2\right)!}\ .
\end{eqnarray*}
For the CP-violating couplings, which only exist in $d=4$, we have
\begin{eqnarray*}
\partial_{t}\tilde{g}_{13} & = & -2\tilde{g}_{13}
-6\kappa\tilde{g}_{13}\\
\partial_{t}\tilde{g}_{14} & = & -2\tilde{g}_{14}
-3\kappa\tilde{g}_{14}\\
\partial_{t}\tilde{g}_{16} & = & -2\tilde{g}_{16}
-\kappa\,d\,\tilde{g}_{16}\\
\partial_{t}\tilde{g}_{17} & = & -2\tilde{g}_{17}
-\kappa\left(\frac{7}{2}\tilde{g}_{17}+\tilde{g}_{19}\right)\\
\partial_{t}\tilde{g}_{19} & = & -2\tilde{g}_{19}
-\kappa\left(\frac{1}{2}\tilde{g}_{17}+3\tilde{g}_{19}\right)\\
\partial_{t}\tilde{g}_{20} & = & -2\tilde{g}_{20}
-\kappa\left( \frac{1}{2}\tilde{g}_{17}-\tilde{g}_{19}\right).
\end{eqnarray*}

This set of beta functions has been obtained in the de Donder gauge, which diagonalizes the Hessian of $S_H$. 
Such beta functions depend on the gauge fixing 
as well as on the particular form of the cutoff kernel $R_k$
and on the parametrization of the quantum field.
The dependence of the beta functions of Cosmological and Newton's constants on these choices 
have been studied in detail in several works regarding the Asymptotic Safety scenario for Quantum Gravity, 
see for instance \cite{cpr2} and most recently \cite{gkl}.
It turns out that the qualitative features of the RG flow
are quite stable.
We believe that, within our truncation, the same should be true
for the flow of the couplings $\tilde{g}_i$.
In fact, some preliminary computations show that the 
non-mixing property in 
the flow of the couplings related to the torsion square monomials is
rather stable.


\section{Flow}

It is immediately obvious that the theory space spanned by the couplings $g_j$ 
contains invariant subspaces.
For example the subspace of the pure torsion invariants $(g_1,g_2,g_3,g_{13},g_{14})$ 
and the subspace of the pure nonmetricity invariant $(g_4,g_5,g_6,g_7,g_8,g_{16})$ are invariant subspaces.
Also, the subspaces of parity even and parity odd terms are invariant.
However, a much stronger statements is true:
the matrix of coefficients of the beta functions can be diagonalized and
the whole 17-dimensional space is a direct sum of one-dimensional invariant subspaces.

The flow is diagonalized when written in terms of the couplings
\begin{eqnarray}
h_1&=&g_1\ ,\quad
h_2=g_2\ ,\quad
h_3=g_3\ ,\quad
h_4=\frac{g_4+g_8}{3}\ ,\quad
\nn\\
h_5&=&g_5+\frac{2(d+1)g_4-4g_8}{d(d+3)}\ ,\quad
h_6=g_6+\frac{2(d+1)g_4-4g_8}{d(d+3)}\ ,\quad
h_7=g_7+\frac{4(d+2)g_8-4g_4}{d(d+3)}\ ,\quad
\nn\\
h_8&=&\frac{2g_8-g_4}{3}\ ,\quad
h_9=-\frac{1}{d+3}g_9\ ,\quad
h_{10}=g_{10}+\frac{1}{d+3}g_9\ ,\quad
h_{11}=g_{11}+\frac{1}{d+3}g_9\ ,\quad
\end{eqnarray}
and for the CP-violating sector in $d=4$
\begin{eqnarray}
h_{12}&=&g_{13}\ ,\quad
h_{13}=g_{14}\ ,\quad
h_{14}=g_{16}\ ,\quad
\nn\\
h_{15}&=&\frac{g_{17}+g_{19}}{3}\ ,\quad
h_{16}=\frac{g_{17}-2g_{19}}{5}\ ,\quad
h_{17}=g_{20}-\frac{g_{17}-2g_{19}}{5}\ ,\quad
\label{couplingsh}
\end{eqnarray}
In terms of these new couplings, the action can be written in the form
\be
S_2=\sum_{j=1}^{17} h_j \int dx\sqrt{g} K_j
\ee
where the operators $K_j$ are given by
\begin{eqnarray}
K_1&=&I_1\ ,\quad
K_2=I_2\ ,\quad
K_3=I_3\ ,\quad
K_4=2I_4+I_8-\frac{4(I_5+I_6+I_7)}{d+3}\ ,\quad
\nn\\
K_5&=&I_5\ ,\quad
K_6=I_6\ ,\quad
K_7=I_7\ ,\quad
\nn\\
K_8&=&-I_4+I_8+\frac{2I_5+2I_6-4I_7}{d}\ ,\quad
K_9=I_{10}+I_{11}-(d+3)I_9\ ,\quad
K_{10}=I_{10}\ ,\quad
K_{11}=I_{11}\ ,\quad
\end{eqnarray}
and in $d=4$ also
\begin{eqnarray}
K_{12}&=&I_{13}\ ,\quad
K_{13}=I_{14}\ ,\quad
K_{14}=I_{16}\ ,\quad
\nn\\
K_{15}&=&2I_{17}+I_{19}\ ,\quad
K_{16}=\frac{5}{3}(I_{17}-I_{19})+I_{20}\ ,\quad
K_{17}=I_{20}\ ,\quad
\label{couplingsh}
\end{eqnarray}
and the operators $I_j$ are listed in (\ref{basisI}).

These couplings have beta functions of the form
\be
\label{betah}
\partial_t \tilde h_j=(-(d-2)+\kappa\lambda_j)\tilde h_j
\ee
with the following coefficients:
\begin{eqnarray}
\lambda_1&=&\frac{d^2-7d-12}{4}\ ,\quad
\lambda_2=\frac{(d+1)(d-4)}{4}\ ,\quad
\lambda_3=\frac{(d+1)(d-4)}{4}\ ,\quad
\lambda_4=\frac{d^2-7d-16}{4}\ ,\quad
\nn\\
\lambda_5&=&\frac{(d+1)(d-4)}{4}\ ,\quad
\lambda_6=\frac{(d+1)(d-4)}{4}\ ,\quad
\lambda_7=\frac{(d+1)(d-4)}{4}\ ,\quad
\nn\\
\lambda_8&=&\frac{d^2-7d-4}{4}\ ,\quad
\lambda_9=\frac{d^2-7d-16}{4}\ ,\quad
\lambda_{10}=\frac{(d+1)(d-4)}{4}\ ,\quad
\lambda_{11}=\frac{(d+1)(d-4)}{4}\ ,\quad
\end{eqnarray}
and furthermore in $d=4$
\begin{eqnarray}
\lambda_{12}=-6\ ,\quad
\lambda_{13}=-3\ ,\quad
\lambda_{14}=-4\ ,\quad
\lambda_{15}=-4\ ,\quad
\lambda_{16}=-\frac{5}{2}\ ,\quad
\lambda_{17}=0\ .\quad
\label{eigh}
\end{eqnarray}

\subsection{Four dimensions}

Let us now consider the special case $d=4$.
In this case the dimensionful couplings $h_2$, $h_3$ $h_5$, $h_6$, $h_7$, $h_{10}$, $h_{11}$
and $h_{17}$ have vanishing beta function.
The corresponding dimensionless couplings have classical scaling, and behave like 
\be
\tilde h_j=\tilde h_{j0}(k/k_0)^{-2}\ .
\ee
They are asymptotically free, and blow up in the infrared.
All the remaining dimensionless couplings have negative $\lambda$.
In terms of the dimensionful couplings
the solution of the flow has the form
\be
G[k]=\frac{G_0}
{1+\frac{1}{2}B G_0 k^2}\ ;\qquad
h_j[k]=h_{j0}
\left(1+\frac{1}{2}B G_0 k^2\right)^{\lambda_j/B\pi}
\ee
In this form one sees that these couplings have finite
limits $G_0$ and $h_{j0}$ for $k\to0$.
On the other hand in the UV limit
$\tilde G\to 2/B$ and, since $\lambda_j/\pi B<0$,
$\tilde h_j\to0$.

\subsection{The Holst subsector}

The Holst action \cite{holst} is defined for an independent
tetrad $\theta$ and a torsionful connection $A$.
Interpreting these fields as one-forms, it can be written 
\be
\label{holst}
\newt\int d^4x
\left[\varepsilon_{abcd }\left(
2\Lambda\,\theta^a\wedge\theta^b\wedge\theta^c\wedge\theta^d
-F^{ab}\wedge\theta^c\wedge\theta^d\right)
+\frac{1}{\gamma}F_{ab}\wedge\theta^a\wedge\theta^b\right]
\ee
where $\gamma$ is the Barbero-Immirzi parameter \cite{immirzi,barbero}.

What sets aside this action
is the fact that it does not require invertibility of the metric.
In fact, the Holst action is well-defined also when the
soldering form (and consequently the metric) becomes degenerate or even zero.
Because of this special property, this action plays a prominent role
in loop quantum gravity.

If the connection is allowed also to be non-metric,
there are (at least) two different ways of generalizing 
the Holst action.
The first is to write the Palatini term for the 
(torsionful, but metric) connection $\mit\Gamma+\beta$,
the second is to write the Palatini term for the
(torsionful and non-metric) connection $\mit\Gamma+\alpha+\beta$.
In both cases the action makes sense also for degenerate tetrads,
but, once rewritten in the form (\ref{act}),
they differ in the coefficients of the $S_2$ terms.

If we define the generalized Holst action by the connection
$\mit\Gamma+\beta$,
we can view it as a special case of (\ref{act}) with
\be
\label{rel1}
-g_2=g_3=\newt\ ;
\qquad
g_{13}=\newt/\gamma\ ;
\qquad
g_i=0,\ \mathrm{for}\ i\not=\{2,3,13\}\ .
\ee

If we define the generalized Holst action by the connection
$\mit\Gamma+\alpha+\beta$,
we can view it as a special case of (\ref{act}) with
\be
-g_2=g_3=g_5=-g_7=-g_9=g_{10}=-g_{11}=\newt\ ,
\quad
g_{13}=2g_{17}=-2g_{19}=2g_{20}=\newt/\gamma\ ,
\ee
the others being zero.
Using the flow equations, we can ask whether the Holst
subclass of actions is closed under the RG flow.
Let us consider the case when the Holst action is defined by the
connection $\mit\Gamma+\beta$.
In this case one sees from equations (\ref{betah},\ref{eigh})
that the beta functions of $g_2=h_2$ and
$g_3=h_3$ are both zero.
This is not the case of the beta function of $\newt$.
Thus relation (\ref{rel1}) is not preserved by the flow.
In the case when the Holst action is defined by the
connection $\mit\Gamma+\alpha+\beta$ the same is true.
Thus already for pure gravity the Holst action is
not stable under the RG flow.

\section{Coupling to matter}

As already mentioned in the introduction, the action of scalars and gauge fields
does not require using a gravitational connection.
The coupling of such fields to torsion and nonmetricity is thus
unnatural at best.
The coupling of fermions to torsion has been considered several
times in the literature, see e.g. (\cite{hehl,shapiroReview}).
Parity-violating effects seem particularly interesting
\cite{mercuri,friedel,perez}.
Spinor fields are by definition representations of the Lorentz group,
and can only couple to metric connections.
If $A$ and $g$ are given, and $A$ is not metric with respect to $g$,
one can still define a coupling of spinors to $A$ by means of the
following construction.
Using the decomposition (\ref{phi},\ref{alphabeta})
one can first extract from $A$ and $g$ the tensors $\alpha$ and $\beta$
and then define the connection $\hat A=\mit\Gamma+\beta$
which is metric by construction.
The covariant derivative acting on a spinor is then defined by:
\begin{eqnarray*}
{\hat{\nabla}}_{\mu}\psi & = & \left(\partial_{\mu}
+\hat A_{\mu}^{\,\,\, ab} \Sigma_{ab}\right)\psi,\qquad \Sigma_{ab}
=\frac{1}{8}\left[\gamma_{a},\gamma_{b}\right].
\end{eqnarray*}
The Lagragian for a free Dirac spinor is
\begin{eqnarray}
\label{spinaction}
S_{1/2} & = & 
\frac{i}{2}\int d^dx \det(e)
\left[\bar{\psi}\gamma^{a}e_{a}^{\mu}{\hat{\nabla}}_{\mu}\psi-{\hat{\nabla}}_{\mu}\bar{\psi}\gamma^{a}e_{a}^{\mu}\psi\right]
\end{eqnarray}
where we took care to define an hermitian action. 
Now we have to recall that integration by parts generates
terms containing torsions and/or nonmetricity.
Namely given two tensors 
$B^{\mu\alpha\cdots}$ and $C_{\alpha\dots}$
and a connection $\nabla$ we have
\begin{eqnarray*}
\int d^dx\sqrt{g}\left(\nabla_{\mu}B^{\mu\alpha\cdots}
C_{\alpha\dots}\right) 
& = & \int d^dx\sqrt{g}\left[-B^{\mu\alpha\cdots}\nabla_{\mu}C_{\alpha\dots}
+\phi_{\mu\,\,\,\,\rho}^{\,\,\,\,\mu}B^{\rho\alpha\cdots}
C_{\alpha\dots}\right]\\
& = & 
\int d^dx\sqrt{g}\left[-B^{\mu\alpha\cdots}\nabla_{\mu}C_{\alpha\dots}
+\left(\frac{1}{2}Q_{\rho\,\,\,\,\mu}^{\,\,\,\,\mu}-T_{\rho\,\,\,\,\mu}^{\,\,\,\,\mu}\right)B^{\rho\alpha\cdots}
C_{\alpha\dots}\right].
\end{eqnarray*} 
Applying this formula to (\ref{spinaction}) we get
\begin{eqnarray*}
S_{1/2} & = & 
=
i\int d^dx\, e\,\bar{\psi}{\cal D}\psi
\end{eqnarray*}
where
\be
{\cal D}=e_{a}^{\mu}{\hat{\nabla}}_{\mu}
+\frac{1}{2} T_{\mu\,\,\,\,\rho}^{\,\,\,\rho}\bar{\psi}\gamma^{a}e_{a}^{\mu}
\ee
is the Dirac operator.
At this point, to calculate the contribution of spinors to the running
of the gravitational couplings, we square the Dirac operator.
A straightforward calculation leads to

\be
{\cal D}^{2}=-\hat\nabla^2+B^{\rho}{\hat{\nabla}}_{\rho}+X
\ee
where
\begin{eqnarray*}
B^{\rho} & = & \frac{1}{4}\left[\gamma^{\mu},\gamma^{\nu}\right] T_{\mu\,\,\,\,\nu}^{\,\,\,\rho} 
- T^{\rho\alpha}{}_\alpha\mathbb{I}
\\
X & = & \frac{1}{4}\hat F_{\mu\nu}{}^{\mu\nu}\mathbb{I} 
- \frac{1}{2}{\hat{\nabla}}_{\mu}T^{\mu\alpha}{}_\alpha \mathbb{I}
-\frac{1}{4}\left[\gamma^{\mu},\gamma^{\nu}\right]\left({\hat{\nabla}}_{\mu}T_{\nu}{}^\alpha{}_\alpha\right)
-\frac{1}{4}T^{\mu\alpha}{}_\alpha T_{\mu}{}^\beta{}_\beta\mathbb{I}.
\end{eqnarray*}
We consider the flow equation equipped with a type two cutoff which
is the correct choice for fermions in the standard case \cite{Dona:2012am}. We have (dropping
surface terms) 
\begin{eqnarray*}
\partial_{t}\Gamma_{k} & = & -\frac{1}{2} \mbox{tr}\left[Q_{d/2-1}\left(\frac{\partial_{t}R_{k}\left({\cal D}^{2}\right)}{{\cal D}^{2}+R_{k}\left({\cal D}^{2}\right)}\right)B_{2}\left({\cal D}^{2}\right)\right]\\
&=& -\frac{1}{(4\pi)^{d/2}}\frac{k^{d-2}}{\left( \frac{d}{2}-1 \right)!} 2^{\left[  \frac{d}{2} \right]} \left[ \frac{1}{16} T_{\alpha\beta\gamma} T^{\alpha\beta\gamma} - \frac{1}{8} T_{\alpha\beta\gamma}  T_{\alpha\gamma\beta} \right] 
\\
&=& -\frac{1}{(4\pi)^{d/2}}\frac{k^{d-2}}{\left( \frac{d}{2}-1 \right)!} 2^{\left[ \frac{d}{2} \right]} 
\left[ \frac{1}{4} \beta_{\alpha\beta\gamma} \beta^{\alpha\beta\gamma} 
- \frac{1}{2} \beta_{\alpha\beta\gamma} \beta^{\alpha\gamma\beta} \right] .
\end{eqnarray*}
In the second step we have used the optimized cutoff
\cite{litim} and the following formula 
for the heat kernel coefficients of the operator $\Delta=-g^{\mu\nu}\hat{\nabla}_{\mu}\hat{\nabla}_{\nu} +B^\mu \hat{\nabla}_{\mu} +X$ \cite{Obukhov,Gusynin}:
\begin{eqnarray*}
b_{2}(\hat{\Delta}) & = & 
\int d^dx\sqrt{g}\mathbb{I}\left[\frac{R}{6}-X+\frac{1}{2}\nabla_{\mu} T_{\quad \alpha}^{\mu \alpha}-\frac{1}{4} T_{\quad \alpha}^{\mu \alpha} T_{\mu \;\, \beta}^{\;\,\, \beta}
-\frac{1}{2} T_\alpha {}^\beta{}_\beta B^\alpha +\frac{1}{2} \nabla_\mu B^\mu -\frac{1}{4} B_\mu B^\mu 
\right].
\end{eqnarray*}

Using the irreducible decomposition for torsion one can check that this only depends on its totally antisymmetric part.
We observe that spinors contribute only to the beta functions of $g_1$ and $g_2$:
\begin{eqnarray}
\partial_{t}\tilde{g}_{1} & = & -\left(d-2\right)\tilde{g}_{1}
+\kappa\frac{1}{4} ((d-7) d-12) \tilde{g}_1
-\frac{1}{(4\pi)^{d/2}}
\frac{1}{\left( \frac{d}{2}-1 \right)!} 2^{\left[ \frac{d}{2} \right]-2} 
\\
\partial_{t}\tilde{g}_{2} & = & -\left(d-2\right)\tilde{g}_{2}+\kappa\frac{1}{4} (d-4) (d+1) \tilde{g}_2
+\frac{1}{(4\pi)^{d/2}}
\frac{1}{\left( \frac{d}{2}-1 \right)!} 2^{\left[ \frac{d}{2} \right]-1} 
\end{eqnarray}
We see that the fermions further contribute to break the special
combination of coefficients (\ref{rel1}) appearing in the Holst action.
Indeed now, even when $g_1=0$ the beta function of $g_1$ is not,
so that a $g_1$ term is generated by the fermions.

A further consequence of the new terms is that the beta functions now
admit new nontrivial fixed points. We shall add the contribution of
one fermion to the beta functions and refer to \cite{Dona:2013qba} for a complete discussion regarding an arbitrary number
of matter fields.
For definiteness we consider a type $\mathbb{I}$a cutoff for the gravitons (see \cite{cpr2} for the nomenclature
regarding the various types of cutoff action).
In $d=4$ we find the fixed points reported in table below.
\begin{center}
\begin{tabular}{ | l | r | r | r | r | r | r |}
\hline
            & $ \tilde{\Lambda} $   & $ \tilde{G} $ & $\tilde{g}_1$ & $\tilde{g}_2$ & $\tilde{g}_3$\\ \hline
FP$_1$ & $0$  & $0$ & $-0.00316629$ & $0.00633257$ & $0$ \\  \hline
FP$_2$ & $-0.3$  & $1.88496$ & $-0.0018591$ & $0.00633257$ & $0$ \\ \hline\hline
& $ \tilde{\Lambda} $   & $ \tilde{G} $ & $1/\tilde{g}_1$ & $1/\tilde{g}_2$ & $1/\tilde{g}_3$ \\ \hline
FP$_3$ & $-0.3$  & $1.88496$ & $-537.893$ & $157.914$ & $0$ \\ \hline
FP$_4$ & $-0.3$  & $1.88496$ & $-537.893$ & $0$ & $0$ \\ \hline
FP$_5$ & $0$  & $0$ & $-315.827$ &  $157.914$ & $0$ \\ \hline
FP$_6$ & $0$  & $0$ & $-315.827$ & $0$ & $0$ \\ \hline
FP$_7$ & $0$  & $0$ & $0$ & $157.914$ & $0$ \\ \hline
FP$_8$ & $-0.3$  & $1.88496$ & $0$ &  $157.914$ & $0$ \\ \hline
FP$_9$ & $-0.3$  & $1.88496$ & $0$ & $0$ & $0$ \\ \hline
FP$_{10}$ & $0$  & $0$ & $0$ & $0$ & $0$ \\ \hline
\end{tabular}
\end{center}
Computing the stability matrix associated to the above fixed points we note that FP$_2$ is UV attractive, i.e.: the stability matrix has five negative eigenvalues, while the other fixed points have a mixture of negative and positive critical exponents.
As we already stressed a fully fledged computation of the RG flow of these couplings in the UV regime should include also curvature squared terms.
Nevertheless we believe that this result strongly hints to the possible existence of non--trivial fixed points in the torsion sector when fermions are present.

\bigskip
{\bf Acknowledgements}. We would like to thank D. Benedetti, M. Reuter
and I. Shapiro for discussions.
C.P. acknowledges the support of the Foundation
Blanceflor Boncompagni Ludovisi, n\'ee Bildt.

\goodbreak

\appendix

\section{Bases of invariants}

Using only the symmetry properties, one can form, in any dimension, eleven different contractions of the tensors $T_{\alpha\beta\gamma}$ and $Q_{\alpha\beta\gamma}$.
In $d=4$ one can construct further nine combinations
using the $\varepsilon$ tensor:
\begin{eqnarray}
\label{basisJ}
&&J_1=T_{\lambda \mu 	\nu}T^{\lambda \mu \nu}~~~
J_2=T_{\lambda \mu \nu}T^{\lambda \nu \mu}~~~
J_3=T_{\: \,\, \mu \lambda}^{\mu}T_{\nu}^{\: \,\, \nu \lambda}
\\
&&J_4=Q_{\lambda \mu \nu}Q^{\lambda \mu \nu}~~~
J_5=Q_{\lambda \mu \nu}Q^{\nu \mu \lambda}~~~
J_6=Q_{\lambda \mu}^{\quad \mu}Q_{\quad \nu}^{\lambda \nu}~~~
J_7=Q_{\;\,  \mu \lambda}^{\mu}Q_{\nu}^{\;\, \nu \lambda}~~~
J_8=Q_{\;\, \mu \lambda}^{\mu}Q_{\,\,\,\,\,\, \nu}^{\lambda \nu}
\nonumber\\
&&J_9=T_{\lambda \mu \nu}Q^{\lambda \mu \nu}~~~
J_{10}=T_{\;\, \mu \lambda}^{\mu}Q_{\quad \nu}^{\lambda \nu}~~~
J_{11}=T_{\;\, \mu \lambda}^{\mu}Q_{\nu}^{\;\, \nu \lambda}
\nonumber\\
&&J_{12}=\varepsilon^{\alpha \beta \gamma \delta }T_{\alpha\beta\eta}T_{\gamma\delta}{}^\eta~~~
J_{13}=\varepsilon^{\alpha \beta \gamma \delta }T_{\alpha\beta\gamma}T_{\delta\eta}{}^\eta~~~
J_{14}=\varepsilon^{\alpha \beta \gamma \delta }T_{\alpha\beta\eta}T_\gamma{}^\eta{}_\delta~~~
J_{15}=\varepsilon^{\alpha \beta \gamma \delta }T_{\alpha\eta\beta}T_\gamma{}^\eta{}_\delta
\nonumber\\
&&J_{16}=\varepsilon^{\alpha \beta \gamma \delta }Q_{\alpha\beta\eta}Q_{\gamma\delta}{}^\eta
\nonumber\\
&&J_{17}=\varepsilon^{\alpha \beta \gamma \delta }Q_{\alpha\eta}{}^\eta T_{\beta\gamma\delta}~~~
J_{18}=\varepsilon^{\alpha \beta \gamma \delta }Q^\eta{}_{\eta\alpha}T_{\beta\gamma\delta}~~~
J_{19}=\varepsilon^{\alpha \beta \gamma \delta }Q_{\alpha\beta\eta}T^\eta{}_{\gamma\delta}~~~
J_{20}=\varepsilon^{\alpha \beta \gamma \delta }Q_{\alpha\beta\eta}T_\gamma{}^\eta{}_\delta
\nonumber
\end{eqnarray}
In terms of these invariants the action can be written
\be
\label{actTQext}
\Gamma_{k}  =  \int d^4x\sqrt{g}\left[\newt\left(2\Lambda-R\right)
+\sum_{j=1}^{20}\bar A_j J_j\right]
\ee
where the coefficients $\bar A_j$ can be arranged into a column vector
$$
\bar A^T=(a_1,a_2,a_3,b_1,b_2,b_3,b_4,b_5,c_1,c_2,c_3,\bar d_1,\bar d_2,\bar d_3,\bar d_4,\bar d_5,\bar d_6,\bar d_7,\bar d_8,\bar d_9)
$$

The identity
\begin{equation}
\label{kronid}
\varepsilon_{ab[c}{}^{[e} \delta_{d]}{}^{f]}=-\varepsilon_{cd[a}{}^{[e} \delta_{b]}{}^{f]} \nonumber
\end{equation}
gives rise to three additional relations between the parity-odd invariants:
\begin{eqnarray}
\nonumber
2J_1+J_2-J_3&=&0\ ,
\\
2J_1-J_2-3J_3-J_4&=&0\ ,
\\
J_6-J_7-2J_8+J_9&=&0\ .
\nonumber
\end{eqnarray}
We can use these relations to eliminate three terms from the action.
We choose to eliminate the terms $d_2 J_2$, $d_3 J_3$ and $d_8 J_8$.
Then the action can be rewritten as
\be
\label{actTQred}
\Gamma_{k}  =  \int d^4x\sqrt{g}\left[\newt\left[2\Lambda-R\right]
+\sum_{j=1}^{17} A_j J_j\right]\ ,
\ee
where the coefficients $A_k$, forming the column vector
$$
A^T=(a_1,a_2,a_3,b_1,b_2,b_3,b_4,b_5,c_1,c_2,c_3,d_1,d_4,d_5,d_6,d_7,d_9)\ ,
$$
are related to the $\bar A_j$ by
\begin{eqnarray}
&&
a_j=\bar a_j\ ;\quad
b_j=\bar b_j\ ;\quad
c_j=\bar c_j\ ;
\nonumber\\
&&
d_1=\bar d_1-\bar d_2+\bar d_3\ ;\quad
d_4=\bar d_4+\frac{1}{4}\bar d_2+\frac{1}{4}\bar d_3\ ;\quad
d_5=\bar d_5\ ;\quad
\label{redruleabcd}
\\
&&
d_6=\bar d_6+\frac{1}{2}\bar d_8\ ;\quad
d_7=\bar d_7-\frac{1}{2}\bar d_8\ ;\quad
d_9=\bar d_9+\frac{1}{2}\bar d_8\ ;\quad
\nonumber
\end{eqnarray}

As explained in the main text, we will work mainly with the basis
formed by the tensors $\alpha_{\lambda\mu\nu}$ and $\beta_{\lambda\mu\nu}$.
Again one can first write an overcomplete set of invariants
\begin{eqnarray}
&&
\label{basisI}
I_1=\beta_{\lambda \mu \nu}\beta^{\lambda \mu \nu}~~~
I_2=\beta_{\lambda \mu \nu}\beta^{\mu \lambda \nu}~~~
I_3=\beta_{\: \lambda \mu}^{\lambda} \beta^\nu{}_\nu{}^\mu
\\
&&
I_4=\alpha_{\lambda \mu \nu}\alpha^{\lambda \mu \nu}~~~
I_5=\alpha_{\lambda \mu \nu}\alpha^{\lambda \nu \mu}~~~
I_6=\alpha_{\;\, \lambda \mu}^{\lambda}\alpha_{\,\,\,\, n}^{\mu \nu}~~~
I_7=\alpha_{\;\, \lambda \mu}^{\lambda}\alpha_{\,\,\,\, n}^{\nu \mu}~~~
I_8=\alpha_{\;\, \lambda \mu}^{\mu}\alpha_{\nu}^{\;\, \lambda \nu}
\nonumber\\
&&
I_9=\alpha_{\lambda \mu \nu}\beta^{\lambda \mu \nu}~~~
I_{10}=\alpha_{\;\, \lambda \mu}^{\lambda} \beta^\nu{}_\nu{}^\mu~~~ 
I_{11}=\alpha_{\;\, \mu \lambda}^{\lambda} \beta^\nu{}_\nu{}^\mu
\nonumber\\
&&
I_{12}=\varepsilon_{\alpha\beta\gamma\delta}\beta_\rho{}^{\alpha \beta} \beta^{\gamma \delta \rho}~~~
I_{13}=\varepsilon_{\alpha\beta\gamma\delta}\beta^{\alpha \beta \rho} \beta^{\gamma \delta}{}_\rho~~~
I_{14}=\varepsilon_{\alpha\beta\gamma\delta}\beta^{\rho \alpha \beta} \beta_\rho{}^{\gamma \delta}  ~~~
I_{15}=\varepsilon_{\alpha\beta\gamma\delta}\beta_{\,\,\, \rho}^{\rho \,\,\,\, \alpha}\beta^{\beta\gamma \delta}  
\nonumber\\
&&
I_{16}=\varepsilon_{\alpha\beta\gamma\delta}\alpha^{\rho \alpha \beta} \alpha_{\rho}^{\,\,\, \gamma \delta}
\nonumber\\
&&
I_{17}=\varepsilon_{\alpha\beta\gamma\delta}\alpha^{\rho \alpha \beta }\beta_{\rho}^{\,\,\, \gamma \delta}~~~
I_{18}=\varepsilon_{\alpha\beta\gamma\delta}\alpha^{\rho \alpha \beta} \beta^{ \gamma \delta}{}_{\rho}~~~
I_{19}=\varepsilon_{\alpha\beta\gamma\delta}\alpha_{\,\,\,\, \rho}^{\rho \alpha }\beta^{\beta \gamma \delta}~~~
I_{20}=\varepsilon_{\alpha\beta\gamma\delta}\alpha_{\,\,\, \rho}^{\rho \,\,\,\, \alpha}\beta^{\beta \gamma \delta}
\nonumber
\end{eqnarray}
in terms of which the action reads
\be
\label{actalphabetaext}
\Gamma_{k}  =  \int d^4x\sqrt{g}\left[\newt\left(2\Lambda-R\right)
+\sum_{j=1}^{20} \bar G_j I_j\right]
\ee
where
$$
\bar G=(\bar g_1,\bar g_2,\bar g_3,\bar g_4,\bar g_5,\bar g_6,\bar g_7,\bar g_8,\bar g_9,\bar g_{10},
\bar g_{11},\bar g_{12},\bar g_{13},\bar g_{14},\bar g_{15},\bar g_{16},\bar g_{17},\bar g_{18},\bar g_{19},\bar g_{20})\ .
$$
This time the identity (\ref{kronid}) leads to the relations
\begin{eqnarray}
3 I_{12}+2I_{13}+I_{14}+I_{15}&=&0\ ,
\nonumber\\
\label{identitiesI}
I_{12}+2I_{13}-I_{15}&=&0\ ,
\\
I_{17}+2I_{18}-I_{19}+I_{20}&=&0\ .
\nonumber
\end{eqnarray}
We can use these relations to eliminate the terms $\bar g_{12}I_{12}$, $\bar g_{15}I_{15}$, and $\bar g_{18}I_{18}$.
Then the action can be rewritten as
\be
\label{actalphabetared}
\Gamma_{k}  =  \int d^4x\sqrt{g}\left[\newt\left[2\Lambda-R\right]
+\sum_{j=1}^{17} G_j I_j\right]
\ee
where the coefficients $G_j$, forming the column vector
$$
G=(g_1,g_2,g_3,g_4,g_5,g_6,g_7,g_8,g_9,g_{10},
g_{11},g_{13},g_{14},g_{16},g_{17},g_{19},g_{20})\ ,
$$
are related to the $\bar G_j$ by
\begin{eqnarray}
&&g_j=\bar g_j\ \mathrm{for}\ j=1\ldots 11
\nonumber\\
&&
g_{13}=\bar g_{13}-\bar g_{12}\ ;\quad
g_{14}=\bar g_{14}-\frac{1}{4}\bar g_{12}-\frac{1}{4}\bar g_{15}\ ;\quad
g_{16}=\bar g_{16}\ ;\quad
\label{redruleg}
\\
&&
g_{17}=\bar g_{17}-\frac{1}{2}\bar g_{18}\ ;\quad
g_{19}=\bar g_{19}+\frac{1}{2}\bar g_{18}\ ;\quad
g_{20}=\bar g_{20}-\frac{1}{2}\bar g_{18}\ ;\quad
\nonumber
\end{eqnarray}

Now we can give the relation between the basis of invariants 
used in eqs.(\ref{actTQext},\ref{actTQred}) and the one used 
in eqs.(\ref{actalphabetaext},\ref{actalphabetared}).
For the overcomplete bases we have
\begin{eqnarray}
\bar g_1&=&2\bar a_1-\bar a_2\ ,\quad
\bar g_2=-2\bar a_1+3\bar a_2\ ,\quad
\bar g_3=\bar a_3
\nn\\
\bar g_4&=&2\bar b_1+\bar b_2\ ,\quad
\bar g_5=2\bar b_1+3\bar b_2\ ,\quad
\bar g_6=4\bar b_3+\bar b_4+2\bar b_5\ ,\quad
\bar g_7=2\bar b_4+2\bar b_5\ ,\quad
\bar g_8=\bar b_4
\nn\\
\bar g_9&=&-\bar c_1\ ,\quad
\bar g_{10}=2\bar c_2+\bar c_3\ ,\quad
\bar g_{11}= \bar c_3
\nn\\
\bar g_{12}&=&2\bar d_1-2\bar d_3\ ,\quad
\bar g_{13}= \bar d_1-2\bar d_3+4\bar d_4\ ,\quad
\bar g_{14}= \bar d_1\ ,\quad
\bar g_{15}=2\bar d_2\ ,\quad
\bar g_{16}= \bar d_5
\nn\\
\bar g_{17}&=&-\bar d_8\ ,\quad
\bar g_{18}= -\bar d_8+2\bar d_9\ ,\quad
\bar g_{19}=2\bar d_7\ ,\quad
\bar g_{20}=4\bar d_6+2\bar d_7
\label{abcd2g}
\end{eqnarray}

The relation between the reduced bases is obtained as follows.
Starting with an action of the form (\ref{actTQred})
one can think of it as an action of the form (\ref{actTQext})
where $\bar d_2=\bar d_3=\bar d_8=0$.
Applying the transformation rule (\ref{abcd2g}) and the reduction rules
(\ref{redruleg}) one obtains a linear transformation of $A$ to $G$.
In the parity-even sector ($k=1\ldots 11$), it is the same
as the transformation of $\bar A$ to $\bar G$:
it is enough to remove the bars on both sides of the relations.
In the parity-odd sector one finds instead
\begin{eqnarray}
g_{13}&=& -d_1+4d_4\ ,\quad
g_{14}= \frac{1}{2}d_1\ ,\quad
g_{16}= d_5
\nn\\
g_{17}&=&-d_9\ ,\quad
g_{19}=2d_7+d_9\ ,\quad
g_{20}=4d_6+2d_7-d_9\nn
\end{eqnarray}

As a check, starting with an action of the form (\ref{actalphabetared})
one can think of it as an action of the form (\ref{actalphabetaext})
where $\bar g_{12}=\bar g_{15}=\bar g_{18}=0$.
Applying the inverse of (\ref{abcd2g}) and the reduction rules
(\ref{redruleabcd}) one obtains the linear transformation of $G$ to $A$,
which happens to be the inverse of the linear transformation from $A$ to $G$.

\goodbreak

\end{document}